\documentclass[superscriptaddress,secnumarabic,amssymb,amsmath,
nobibnotes,aps,prd,showkeys,showpacs,nofootinbib,twocolumn]{revtex4}

\newcommand{\bee}{\begin{equation}}
\newcommand{\eee}{\end{equation}}
\newcommand{\eaa}{\end{eqnarray}}
\newcommand{\baa}{\begin{eqnarray}}

\usepackage{graphicx}
\usepackage{bm}
\usepackage{amsmath}
\usepackage{amsfonts}
\usepackage{amssymb}
\usepackage{epstopdf}
\usepackage{natbib}%
\usepackage{color}
\def\ni{\noindent}

\begin{document}

\title{\Large Entropic gravity from noncommutative black holes}

\author{Rafael C. Nunes}
\email{rcnunes@fisica.ufjf.br} \affiliation{Departamento de
F\'isica, Universidade Federal de Juiz de Fora, 36036-330, Juiz de
Fora, MG, Brazil}
\author{Hooman Moradpour}
\email{h.moradpour@riaam.ac.ir} \affiliation{Research Institute
for Astronomy and Astrophysics of Maragha (RIAAM), P.O. Box
55134-441, Maragha, Iran}
\author{Ed\'esio M. Barboza Jr.}
\email{edesiobarboza@uern.br} \affiliation{Departamento de F\'isica, 
  Universidade do Estado do Rio Grande do Norte, 59610-210, Mossor\'o - RN, 
Brasil}
\author{Everton M. C. Abreu}
\email{evertonabreu@ufrrj.br} \affiliation{Departamento de
F\'isica, Universidade Federal de Juiz de Fora, 36036-330, Juiz de
Fora, MG, Brazil} \affiliation{Grupo de F\'isica Te\'orica e Matem\'atica 
F\'isica, Departamento de F\'isica, Universidade Federal Rural do
Rio de Janeiro, 23890-971 Serop\'edica, Rio de Janeiro, Brazil}
\author{Jorge Ananias Neto}
\email{jorge@fisica.ufjf.br} \affiliation{Departamento de
F\'isica, Universidade Federal de Juiz de Fora, 36036-330, Juiz de
Fora, MG, Brazil}

\keywords{black holes, noncommutative space-time, entropic gravity} \pacs{ }
%%%%%%%%%%%%%%%%%%%%%%%%%%%%%%%%%%%%%%%%%%%%%%%%%%%%%%%%%%%%%%%%%%%%%%%%%%%%%

%%%%%%%%%%%%%%%%%%%%%%%%%%%%%%%%%%%%%%%%%%%%%%%%%%%%%%%%%%%%%%%%%%%%%%%%%%%%%
\begin{abstract}
\noindent
In this paper we will investigate the effects of a noncommutative (NC) space-time on the 
dynamics of the universe. We will generalize the black hole entropy formula for a 
NC black hole. Then, using the entropic gravity formalism, we will 
show that the noncommutativity changes the strength of the gravitational 
field. By applying this result to a homogeneous and isotropic universe 
containing nonrelativistic matter and a cosmological constant, we will show that the 
model modified by the noncommutativity of the space-time is a better fit to the obtained 
data than the standard one.
\end{abstract}
%%%%%%%%%%%%%%%%%%%%%%%%%%%%%%%%%%%%%%%%%%%%%%%%%%%%%%%%%%%%%%%%%%%%%%%%%%%%%

\maketitle

\section{Introduction}
\label{intro}

% The general relativity ``nothing-escapes" view of a black hole (BH)
% is changed when quantum effects are taken into account \cite{hawking}.
% The Hawking radiation arises from the coupling between gravity and quantum 
% mechanics. The BH radiation can be understood as the quantum tunneling of 
% vacuum fluctuations through the horizon.

In the last $40$ years, detailed investigation of
several aspects concerning the black hole (BH) physics were carried out. The 
thermodynamics of BHs has been one of most active research lines
in theoretical physics. The pioneering work of Bekenstein \cite{bekenstein}
has shown that the BH entropy is proportional to its
surface area. Thereafter, analyzing the origin of BH's entropy 
following the quantum mechanics, Hawking \cite{hawking} showed
that the BH has a thermal radiation with a
temperature $T_H = \kappa / 2\pi$ ($\kappa=$ surface gravity). Hawking also 
suggested that the majority
of the information concerning the initial states is protected behind
the event horizon. The information will not be back to the
asymptotic region far from the evaporating BH \cite{hawking2}.
From these initial works, several discussions on the
thermodynamics of BHs \cite{varios2} have arisen.

However, in spite of the progress that the theoretical physics has
accomplished in the understanding of this issue, a complete
explanation for the final state of a BH after the evaporation remains unknown.  
The final piece of the puzzle should be a quantum gravity theory (QGT). One of 
the main candidates for QGT is the string theory. It is known that when we 
have a magnetic background in string theory, the
resulting algebra acquires noncommutative (NC) features. Due
to string-BH correspondence principle \cite{susskind} and some
other results concerning noncommutativity
\cite{reviewsNC}, there is a whole literature approaching the
so-called noncommutative black holes (NCBH) \cite{NCBH}.

A powerful way to describe NCBHs is through the so-called
generalized uncertainty principle (GUP) \cite{gup}. When quantum gravitational effects are 
taken into account the usual Heisenberg uncertainty principle is modified  
\cite{varios3}. Thus GUP provides
a minimal length scale and therefore change the thermodynamics of
a singular BH at the Planck scale. The GUP imply yet in the nonvanishing 
of the position coordinates commutator. This noncommutative property is 
described by a NC parameter, which lies at the Planck scale. The NCBHs that 
emerges from the noncommutativity of the space-time at small scales \cite{ss} 
are analogous to
a nonsingular BH with two horizons \cite{dymnikova}.

%Initially, NC space-times were introduced to avoid divergences in 
%the relativistic quantum field theory \cite{snyder}.

Alternatively, another
way to deal with a BH is through the holographic principle, which
merges the effects of quantum mechanics and gravity
\cite{thooft}.

In this work we will use the entropic gravity formalism, in which  
gravitation is obtained from thermodynamic properties of a holographic
surface \cite{Ver} to obtain the modifications on the gravitational 
field from the thermodynamics
of NCBHs. From the NC corrections we will obtain a modified Friedmann equation and 
the we will constrain the NC parameter to some of the most recent observational 
data.

This paper is organized as follows: in Section 2 we will discuss briefly the thermodynamics of NCBHs;
in Section 3 we will present the NC corrections on the gravitational 
field following the entropic gravity formalism and derive 
the equations for Friedmann-Robertson-Walker universe; in Section 4 
we will investigate the feasibility of the model connected to the latest 
observational data; in Section 5 we will present our conclusions.
In this paper we will use the natural unit system: $G = \hbar = c = k_B = 1$. As 
usual, we  will adopt the convention that a $0$ subscript attached to any quantity 
means that it must be evaluated at present time.

\section{Noncommutative black hole thermodynamics in a nutshell}

In a NC space-time, the commutator between the position coordinates is given by

\begin{eqnarray}
[x^\mu, x^\nu]=i\theta^{\mu\nu},
\end{eqnarray}
where $\theta^{\mu\nu}$ is the so-called NC parameter. In the string 
theory, this commutator shows that the coordinates of the target space-time 
become NC operators on a D-brane. The product of two fields on this NC 
space-time is replaced by the Moyal product of commutative fields \cite{moyal}, 
given by
\begin{eqnarray}
&&\!\!\!\!\!\!\!\!\!\!\!\!\!\!\!f(x)\star g(x) \nonumber \\
&=&\bigglb\{\exp 
\left[\frac{i}{2}\theta^{\mu\nu}\frac{\partial}{\partial\alpha^\mu}\frac{
\partial}{\partial\beta^\nu}\right]
f(x+\alpha)g(x+\beta)\biggrb\}\bigglb| 
_{\alpha=\beta=0}\nonumber\\
&=&f(x)g(x) + 
\frac{i}{2}\theta^{\mu\nu}\partial _\mu f(x)\partial_\nu g(x)\nonumber\\
&+& \mathcal{O}(\theta^2).
\end{eqnarray}

The original commutative metric for a Schwarzchild BH solution is
\begin{eqnarray}
ds^2&=&-\left(1-\frac{2M}{r}\right) dt^2+ \left(1-\frac{2M}{r}\right)^{-1} dr^2 
\nonumber\\
&+& \,\, r^2(d\theta^2+\sin^2\theta d\phi^2),
\end{eqnarray}

\ni where $M$ is the BH mass. In the vierbeins representation, the metric tensor
$g_{\mu\nu}$ can be written as
\begin{eqnarray}
\label{vie}
g _{\mu\nu}=e^a _\mu\, e^b _\nu \eta _{ab}.
\end{eqnarray}

\ni Thus, a NCBH solution can be obtained by redefining Eq.(\ref{vie}) as \cite{ss}:
\begin{eqnarray}
\label{gnc}
\tilde{g}_{\mu\nu}=e^a  {_{(\mu}}\star e^b {_{\nu)}} \eta _{ab}.
\end{eqnarray}

\ni In what follows we will consider the case in which the non-vanishing 
components of $\theta^{\mu\nu}$ are $\theta^{23}=\beta$ and 
$\theta^{32}=-\beta$.
In this case, the non-vanishing components of the metric tensor (\ref{gnc}) 
are given by
\begin{eqnarray}
\label{gco}
\tilde{g}_{00}&=&g_{00},\nonumber\\
\tilde{g}_{11}&=&g_{11}+\frac{1}{4}\beta^2 g_{11} \cos (2\theta),\nonumber\\
\tilde{g}_{22}&=&g_{22}-\frac{1}{4}\beta^2 g_{22} \cos (2\theta),\\
\tilde{g}_{33}&=&g_{33}+\frac{1}{8}\beta^2\,\frac{\partial^2g_{33}}{\partial 
\theta^2}, \nonumber\\
\tilde{g}_{12}&=&-\frac{\beta^2}{4}\sqrt{g_{11}g_{22}}\,\sin(2\theta).\nonumber
\end{eqnarray}

\ni From Eqs. (\ref{gco}), it is possible to show that the entropy of a 
NC Schwarzchild BH is given by \cite{ss}
\begin{eqnarray}
\label{ncentropy}
\tilde{S}(r_+)=\left(1-\frac{\beta^2}{4}\right) S(r_+),
\end{eqnarray}
where $S(r_+)=A/4$ is the entropy of the commutative Schwarzchild BH and 
$\tilde{S}(r_+)$ is the entropy of the NC Schwarzchild BH.

%%%%%%%%%%%%%%%%%%%%%%%%%%%%%%%%%%%%%%%%%%%%%%%%%%%%%%%%%%%%%%%%%%%%%%%%%%%%%%%%%%%%%%%%%%%%
%%%%%%%%%%%%%%%%%%%%%%%%%%%%%%%%%%%%%%%%%%%%%%%%%%%%%%%%%%%%%%%%%%%%%%%%%%%%%%%%%%%%%%%%%%%%

\section{Noncommutative entropic gravity}

According to Verlinde's hypothesis \cite{Ver}, the tendency of any
system to increase their entropy (the second law of
thermodynamics) is the origin of gravity and leads to the
emergence of space-time. In fact, this approach helps us to provide
a thermodynamic description of the gravitational field equations
in various theories
\cite{Cai4,Smolin,Li,Tian,Myung1,Vancea,Modesto,Sheykhi1,BLi,Sheykhi2,
  Sheykhi21,Sheykhi22,Sheykhi23,Ling,Sheykhi24,Gu,Miao1,other,mann,SMR,ms}.
Since the entropy formula plays a key role in this approach, any
modification of the system entropy may affect the gravitational
field equations and therefore the corresponding Friedmann equations
\cite{Sheykhi2,Sheykhi21,Sheykhi23,Sheykhi24}. Moreover, it is
worthwhile mentioning here that Verlinde's interpretation of the
origin of gravity and space-time is in line with both the generalized
entropy formula and its corresponding cosmology \cite{m}.

In order to use Verlinde's approach, we need to evaluate the
entropy of the system. For this purpose, consider a system of energy $E$ enclosed 
by the
surface $A$. By generalizing Eq. \eqref{ncentropy} to the system boundary, a
surface of radii $r_h=2M$ assumed to be the holographic screen \cite{Ver}, the 
entropy of the system is given by
\begin{equation}\label{ent1}
S_{A} = \Big(1-\frac{\beta^2}{4} \Big)\frac{A}{4},
\end{equation}

\ni where $A=4\pi r_h^2$ is the surface
area of the boundary and $\beta=const.$ is the NC parameter. Since the boundary 
surface consists of $N$ degrees of
freedom, we can write \cite{pad}
\begin{eqnarray}\label{7}
A=QN,
\end{eqnarray}

\ni where $Q$ is a constant proportional to the square
of Planck length $\ell_p$. %It is useful to note here that in our units 
%$\ell_p=1$. 
According the energy equipartition theorem, the source energy content is 
distributed on the surface degrees of
freedom as \cite{pad}
\begin{eqnarray}\label{8}
E=\frac{1}{2}NT,
\end{eqnarray}

\ni where $T$ is the surface temperature.
Combining Eqs.~(\ref{7}) and~(\ref{8}) we have that
\begin{eqnarray}
\label{9}
T=\frac{QM}{2\pi r_h^2},
\end{eqnarray}

\ni where $M=E$ is the gravitational mass of the source. According to Verlinde's 
approach \cite{Ver}, this tendency of the source to increase its
entropy implies that the force acting on a test particle of mass $m$ is given by
\begin{eqnarray}\label{10}
F\Delta x=-T\Delta S_A,
\end{eqnarray}

\ni where $\Delta x$ is the displacement of the test particle from the 
holographic screen $A$ and the minus sign is due the inward nature of the
entropy flux \cite{SMR}.
If the distance between the test particle and the holographic surface 
is of the order of magnitude of its Compton wavelength 
$\lambda_m=2\pi/m$, the
particle is absorbed by the holographic screen leading to an increase of the
system entropy \cite{Ver}. In this case, we can set
$\Delta x=\eta\lambda_m$, with $\eta\sim 1$. Finally, from (\ref{ent1}), 
(\ref{7}), (\ref{9}) and (\ref{10}) we obtain
\begin{eqnarray}\label{12}
  F&=&-T\frac{\Delta A}{\Delta x}\frac{\Delta S_A}{\Delta A}\nonumber\\
 &=&-\frac{Q^2}{16\pi^2\eta}(1-\frac{\beta^2}{4})\Big(\frac{mM}{r_h^2}\Big),
\end{eqnarray}

\ni where we have used the fact that $\Delta A=A/N=Q$ \cite{Sheykhi21}. The 
Newtonian limit, $F\to -mM/R^2$, obtainable when
$\beta\rightarrow0$, yields $Q=4\pi\eta^{1/2}$. Thus, the gravitational force 

\ni acting on a particle of mass $m$ in the holographic screen is
\begin{eqnarray}\label{NF_NC}
  F=-(1-\frac{\beta^2}{4})\Big(\frac{mM}{r_h^2}\Big).
\end{eqnarray}

\ni and, as the gravitational force is attractive, this fact constrains the NC parameter to have values in 
the range $-2\le\beta\le2$. 
%This means 
%that the noncommutativity can change not only the strength of the gravitational 
%field but also its attractive nature. 
%Thus, in the scenario considered in this paper, the effects of 
%noncommutativity on the gravitational field can be obtained performing the 
%prescription $G_N\to(1-\beta^2/4)G_N$ in the 

The gravitational potential energy and the kinetic energy of the test particle 
are, respectively,
\begin{equation}
  \label{Pot_E}
  U=-(1-\frac{\beta^2}{4})\,\frac{mM}{r_h}
\end{equation}

\ni and
\begin{equation}
 \label{Kin_E}
 K=\frac{1}{2}m\dot{r}^2_h.
\end{equation}

\ni Now, let us apply the above results for an homogeneous and 
isotropic expanding universe. In this case, the 
radius $r_h$ can be written in terms of the comoving distance $x$ as $r_h=a(t)x$, 
where $a(t)$ is the scale factor. Writing the source mass as 
$M=4\pi\rho a(t)^3\,x^3/3$, the energy conservation for the test particle can 
be written as
\begin{equation}
\label{Conserv_E}
E=\frac{1}{2}m\dot{a}^2x^2-\frac{4\pi}{3}(1-\frac{\beta^2}{4})\,m\rho\,a^2
x^2.
\end{equation}

\ni Multiplying both sides by $2/ma^2x^2$ and rearranging the terms we can write that
\begin{equation}
  \label{Friedmann_eq}
  H^2=\frac{8\pi}{3}(1-\frac{\beta^2}{4})\rho-\frac{\kappa}{a^2},
\end{equation}

\ni where $H=\dot{a}/a$ is the Hubble parameter and $\kappa=-2E/mx^2$. As we 
can see, the above equation is a modified version of the Friedmann equation, 
which is obtained in the limit $\beta\to0$. Thus, 
in the entropic gravity formalism, the NC of the space-time affects the way 
the universe evolves. 

Since the energy stored in the source is $U=M=4\pi\rho a^3x^3$, the first law 
of thermodynamics reads as:
\begin{eqnarray}
 \label{Therm_FL}
 TdS&=&dU+pdV\nonumber\\
 &=&4\pi a^2 x^3\Big[\frac{1}{3}ad\rho+(\rho+p)da\Big].
\end{eqnarray}

\ni Assuming a reversible expansion, $dS=0$, is easy to show that 
\begin{equation}
 \label{fluid_eq}
 \dot{\rho}+3H(\rho+p)=0.
\end{equation}

\ni In the above equations, $p$ is the pressure of the fluid. By combining 
(\ref{Friedmann_eq}) and (\ref{fluid_eq}) we obtain the acceleration 
equation as being
\begin{equation}
\label{acc_eq}
\frac{\ddot{a}}{a}=-\frac{4\pi}{3}(1-\frac{\beta^2}{4})(\rho+3p).
\end{equation}

\ni Summarizing, the final effect of the NC entropic cosmology is the modification of the 
Newton constant, $G_N$ to $(1-\beta^2/4)G_N$ in the usual field equations.

\section{Entropic cosmology with noncommutative effects}

\begin{figure*}[t!]
\includegraphics[width=6in, height=5.0in]{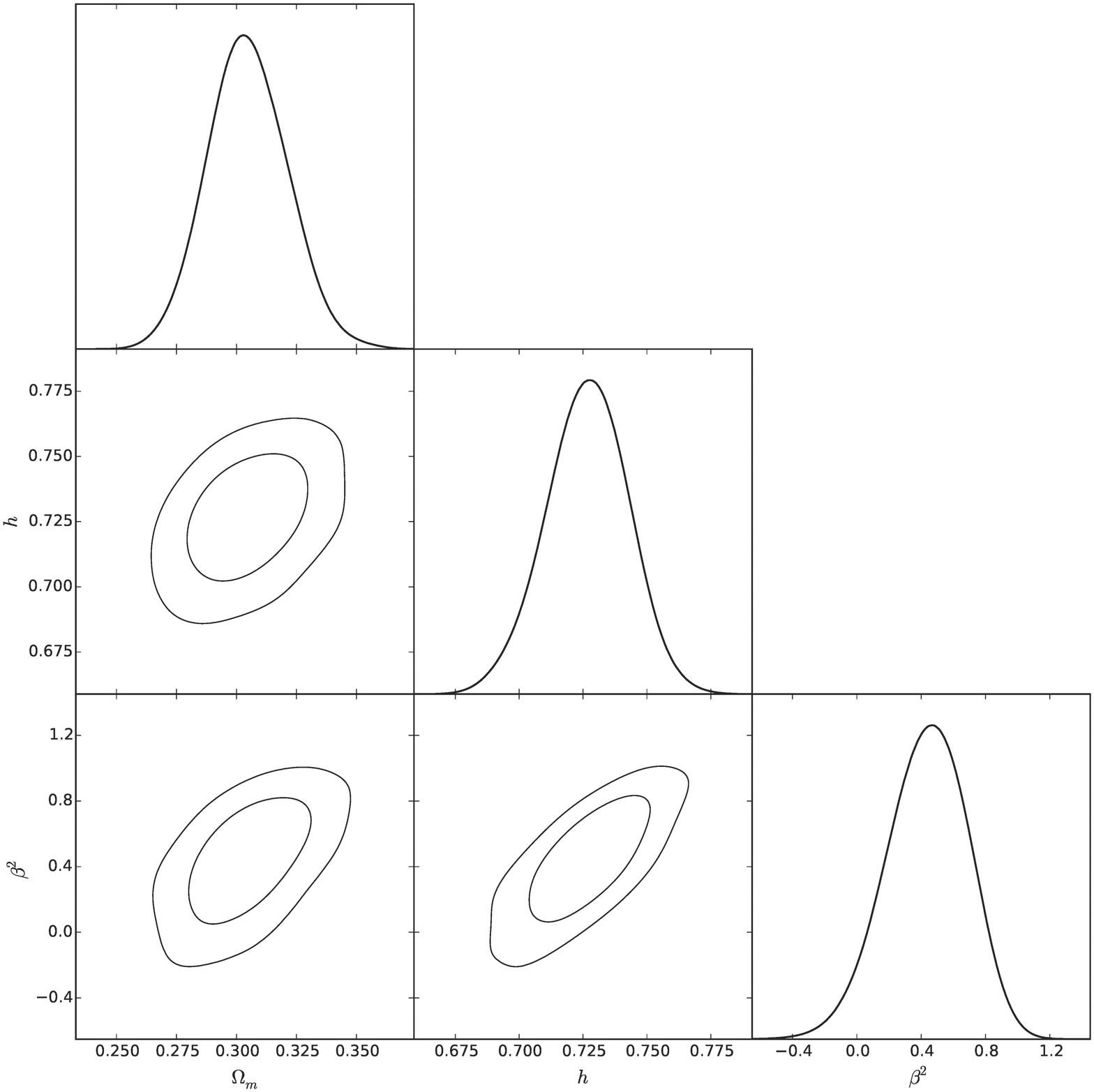}
\caption{\label{fig1} 1$\sigma$ and 2$\sigma$ confidence contours for 
the model parameters obtained from a joint analysis with JLA, BAO, CC and $H_0$ 
data sets.}% in the plan $h - \beta^2$.}
\end{figure*}

In order to probe the viability of the cosmological scenario developed in the 
previous section we will consider a spatially flat Friedmann-Robertson-Walker 
universe dominated by a pressureless matter (baryonic plus dark matter) as well as the 
energy of the quantum vacuum 
($p=-\rho_{\Lambda}$). Concerning these considerations the Eq. 
(\ref{Friedmann_eq}) can be rewritten as
\begin{eqnarray}
\label{H_function}
H^2&=&H^2_0 \Big(1-\frac{\beta^2}{4} \Big) \Big[  \Omega_{m,0}(1+z)^3 + 
\Omega_{\Lambda,0} \Big],
\end{eqnarray}

\ni where $\Omega_{i,0}$ denotes the current fractional densities of radiation, 
non-relativistic matter and quantum vacuum, respectively. 
%Motivated by recent and accurate measures on the cosmic microwave background  
% by Planck satellite, without loss of generality, 
% we are here considering $\Omega_k=0$. 
In this case, the normalization condition reads 
% Taking the condition 
% $H(z=0)=H_0$, from the above equantion we find the density normalization 
% condition on the form
\begin{eqnarray}
\label{normalization}
\frac{4}{4 - \beta^2} = \Omega_{\Lambda,0} + \Omega_{m,0};\quad\beta\ne\pm2.
\end{eqnarray}

%In the absence of the NC corrections in eq. (\ref{H_function}) and 
%(\ref{normalization}), i.e. 
%taking $\beta = $ 0, we obtain 
Note that in the absence of the NC corrections, the standard $\Lambda$CDM model 
is recovered, as expected.

In what follows, we will perform an observational analysis of the above 
model. In order to constrain the model parameters we use 
some cosmological probes
that map the late-time universe expansion history.  The data used in our 
analysis are: the ``joint light curves" (JLA) sample 
\cite{snia3} which comprises 740 type Ia supernovae in the redshift range 
$z \in[0.01, 1.30]$; the baryon acoustic oscillation (BAO) measurements 
from the  Six  Degree  Field  Galaxy  Survey  (6dF) \cite{bao1}, 
the  Main  Galaxy  Sample  of  Data  Release 7  of  Sloan  Digital  Sky  Survey 
(SDSS-MGS) \cite{bao2}, the  LOWZ  and  CMASS  galaxy  samples  of  the Baryon  
Oscillation  Spectroscopic  Survey  (BOSS-LOWZ  and  BOSS-CMASS, respectively) 
\cite{bao3} and the distribution of the LymanForest in BOSS (BOSS-Ly) 
\cite{bao4}; the cosmic chronometers (CC) data set, which contains 
30 measurements of $H(z)$ covering the redshift range $0 < z < 2$ \cite{cc}; 
and the recent measurement of the Hubble constant, 
$H_0=73.24 \pm 1.74{\rm km}\cdot{\rm s}^{-1}{\rm Mpc}^{-1}$  \cite{riess}.

We will use CLASS \cite{class} and Monte Python \cite{monte} codes to perform the 
statistical analysis of the model for the combined data set: JLA + BAO + CC + 
$H_0$. 
In order to obtain correlated Markov Chain Monte
Carlo samples from CLASS/Monte Python code, we use the Metropolis Hastings 
algorithm with uniform priors on the model parameters.

\begin{table}[h!]
\begin{center}
\begin{tabular} {l l }
\hline
Parameter & best fit $\pm$ 1$\sigma$  \\ \hline 
$\beta^2$ &$0.3922_{-0.67}^{+0.67}$    \\ 
$h$ &$0.728_{-0.040}^{+0.042}$    \\ 
$\Omega_{m}$ &$0.3034_{-0.043}^{+0.045}$ \\ 
$\Omega_{\Lambda}$ &$0.6966_{-0.046}^{+0.043}$ \\ 
\hline 
\end{tabular}
\end{center}
\caption{\label{tab1} Constraints on the free parameters of the model from 
the combined JLA 
+ BAO + CC + $H_0$ data set.}
\label{tab2}
\end{table} 

\begin{figure}[t!]
\includegraphics[width=3in, height=2.5in]{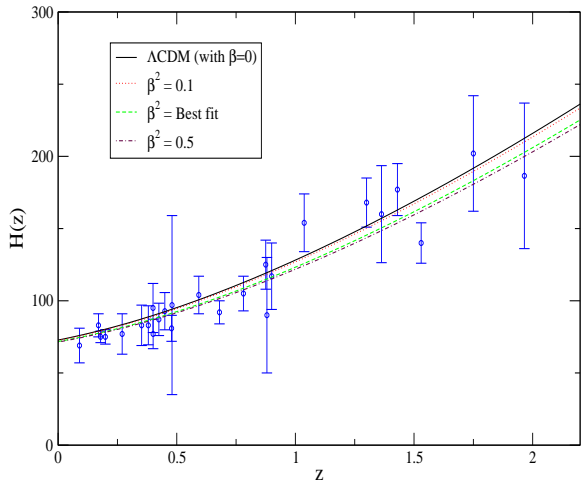}
\caption{\label{fig2} Evolution of the function $H(z)$ in the units of km 
s${}^{-1}$ Mpc${}^{-1}$ for some values of $\beta^2$.}
\end{figure}

Table \ref{tab1} summarizes the main results of our statistical analysis. 
Figure \ref{fig1} shows the parametric spaces $h - \beta^2$ 
($h=H_0/100{\rm km}\cdot{\rm s}^{-1}\cdot{\rm Mpc}^{-1}$), 
$\Omega_{m,0}-\beta^2$ and $\Omega_{m,0}-h$.
The results of our analysis shows that, at 
1$\sigma$ confidence level, the modified $\Lambda$CDM model fits the 
data better than the standard $\Lambda$CDM model, i.e., without the NC 
effects. However, the standard $\Lambda$CDM model still remains in good 
agreement with the data since $-0.08 \leq \beta^2 \leq 0.92$ 
at 2$\sigma$ confidence level. 
Therefore, it's characterized, at least for the data combination used in this 
paper, that NC effects, encoded in the parameter $\beta $,  
can not be ruled out. Figure 
\ref{fig2} shows the effects of the parameter 
$\beta^2$ on the expansion rate of the universe front to the CC data.
\\

\section{Final remarks}

To explain the current accelerated phase of the universe is one of the main 
challenges of the theoretical physics nowadays. The knowledge acquired in many 
areas of physics has been put together and it was applied to cosmology in the hope 
that a better understanding of the physics behind the cosmic acceleration 
can be achieved. 

Considering this line of research, we have investigated the implications of a 
NC geometry to the late time expansion of the universe from the 
entropic gravity perspective. In the entropic gravity formalism developed 
by Verlinde \cite{Ver}, the origin of the gravitational field is associated 
with the perturbations in the information manifold due to particles' 
motion connected to the holographic screen. Since the noncommutativity of the 
space-time changes the value of the black hole entropy, the gravitational field emerging 
from the entropic principle will be different from the usual Newtonian 
gravitational filed. 

In this paper we have investigated the consequences of 
this modification to the late time expansion of the universe. The correspondent 
NC correction of the Friedmann equation is measured by the NC parameter 
$\beta$. Assuming flatness and that the universe contains only 
nonrelativistic matter and a cosmological constant, we constrain the NC 
parameter with the latest observational data of type Ia supernovae 
distance, baryon acoustic oscillations and cosmic chronometers. Our results 
shows that the NC corrected $\Lambda$CDM model is favored compared with the 
standard one. The results obtained here are in agreement with the results 
obtained from the entropic gravity formalism in the framework of nongaussian 
statistics which also favor a weaker gravitational field 
\cite{nossos}.

\acknowledgments

The work of H. Moradpour has been supported financially by
Research Institute for Astronomy \& Astrophysics of Maragha
(RIAAM) under project No.1/4165-10. The Brazilian authors thank CNPq
(Conselho Nacional de Desenvolvimento Cient\' ifico e
Tecnol\'ogico), Brazilian scientific support federal agency, for
partial financial support, Grants numbers 302155/2015-5,
302156/2015-1 and 442369/2014-0 and E.M.C. Abreu thanks the
hospitality of Theoretical Physics Department at Federal
University of Rio de Janeiro (UFRJ), where part of this work was
carried out.

\noindent

%%%%%%%%%%%%%%%%%%%%%%%%%%%%%%%%%%%%%%%%%%%%%%%%%%%%%%%%%%%%%%%%%%%%%%%%%%%%%%%%%%%%%%%%

\end{document}